\documentclass[12pt, preprint]{aastex}
\begin{document}
\title{MODELING THE ALIGNMENT PROFILE OF SATELLITE GALAXIES IN CLUSTERS}
\author{Hyunmi Song and Jounghun Lee}
\affil{Astronomy Program, FPRD, Department of Physics and Astronomy,
Seoul National University, Seoul 151-747, Korea\\
\email{hmsong@astro.snu.ac.kr, jounghun@astro.snu.ac.kr}}
\begin{abstract}
Analyzing the halo and galaxy catalogs from the Millennium simulations at 
redshifts $z=0,\ 0.5,\ 1$, we determine the alignment profiles of cluster 
galaxies by measuring the average alignments between the major axes of 
the pseudo inertia tensors from all satellites within cluster's virial radius 
and from only those satellites within some smaller radius as a function of 
the top-hat scale difference.  The alignment profiles quantify how well the 
satellite galaxies retain the memory of the external tidal fields after merging 
into their host clusters and how fast they lose the initial alignment 
tendency as the cluster's relaxation proceeds.
It is found that the alignment profile drops faster at higher redshifts and on 
smaller mass scales.
This result is consistent with the picture that the faster merging of the 
satellites and earlier onset of the nonlinear effect inside clusters tend to 
break the preferential alignments of the satellites with the external tidal 
fields. Modeling the alignment profile of cluster galaxies as a power-law of the 
density correlation coefficient that is independent of the power spectrum 
normalization ($\sigma_{8}$) and demonstrating that the density 
correlation coefficient varies sensitively with the density parameter 
($\Omega_{m}$) and neutrino mass fraction ($f_{\nu}$), we suggest that the 
alignment profile of cluster galaxies might be useful for breaking 
the $\Omega_{m}$-$\sigma_{8}$ and $f_{\nu}$-$\sigma_{8}$ degeneracies.  
\end{abstract}
\section{INTRODUCTION}

The most salient feature of the large scale structure in the Universe is the 
web-like filamentary pattern of the spatial distribution of the galaxies, which 
is often referred to as the cosmic web \citep{web96}. Ever since the work of 
\citet{HP75} who proposed testing the statistical significance of the 
anisotropy in the orientations of the galaxies embedded in the large-scale 
structure, plenty of observational and numerical researches have been conducted 
to understand the origin, property and dynamics of the cosmic web within the 
basic framework of the standard $\Lambda$CDM ($\Lambda$+cold dark matter) 
cosmology \citep[e.g.,][]
{west89,west-etal91,schmalzing-etal99,erdogdu-etal04,colberg-etal05,dolag-etal06,
shen-etal06,faucher-etal08,forero-etal09,zhang-etal09,shandarin-etal10,
gay-etal10,aragon-etal10a,aragon-etal10b,bond-etal10,murphy-etal11,NC11,
shandarin11,sousbie11,veraciro-etal11}. 

It is now understood that the anisotropic spatial correlations of the 
gravitational tidal fields mould the large scale structure into the cosmic web 
\citep[e.g.,][]{altay-etal06,wang-etal11}, which interconnects the 
Universe on different scales and translates the large-scale anisotropy into 
the small-scale intrinsic alignments 
\citep[e.g.,][]{binggeli82,west-etal91,west-etal95,web96,
colberg-etal99,pen-etal00,WB00,PB02,knebe-etal04,navarro-etal04,zenter-etal05,
faltenbacher-etal05,lee-etal05,BS05,colberg-etal05,KE05,hahn-etal07,LE07,
faltenbacher-etal08,LHP09,faltenbacher-etal09, nie-etal10,paz-etal11,
wang-etal11,lee11}. 

Many observational phenomena can attribute to the interconnection between 
the large-scale anisotropy and the small-scale intrinsic alignments. 
For instance, \citet{GF10} reported the detection of strong correlations 
between the orientations of the galaxy groups and the neighbor group 
distributions on the scale up to $20\,h^{-1}$Mpc. 
\citet{paz-etal11} found a strong alignment signal between the projected major 
axes of the group shapes and the surrounding galaxy distribution up to scales of 
$30\,h^{-1}$Mpc. They noted that this observed anisotropy signal becomes 
larger as the galaxy group mass increases, which is in excellent agreement with 
the prediction of the $\Lambda$CDM cosmology. Their result is also consistent 
with the previous findings that the degree of the alignment of the cluster 
galaxies depends on the velocity dispersion and number of the cluster galaxies 
\citep{plionis-etal03,aryal-etal07,godlowski-etal10,godlowski11b}. 
\citet{smargon-etal11} detected a weak signal of the alignments between the 
projected major axes of the neighbor clusters on the scale of $20 h^{-1}$Mpc 
as well as a strong signal of the alignments between the cluster major axis and 
the direction to the other clusters on scales up to $100\,h^{-1}$Mpc.
\citep[see also][]{hashimoto-etal08,wang-etal09,hao-etal11,SM11}. 
\citet{trujillo-etal06} and \citet{varela-etal12} found that the spin axes 
of the spiral galaxies in the vicinity of the voids tend to lie on the surfaces 
of the voids, while \citet{jones-etal10} showed that the spin axes of the spiral 
galaxies located in the cosmic filaments tend to be aligned with the elongated 
axes of the host filaments. For a comprehensive review on the issue of the 
galaxy orientation and its relation to the structure formation, 
see \citet{schafer09} and \citet{godlowsk11a}. 
 
According to the previous works, the nonlinear evolution
of the gravitational tidal fields has dual effects. On the one hand it sharpens 
the cosmic web and enhances the degree of the large-scale anisotropy by inducing 
filamentary merging and accretion. On the other hand it diminishes the 
interconnections between small and large scales by developing secondary filaments 
and local effects. 
To explain the formation and evolution of the large-scale structure in the 
cosmic web, it is first necessary to understand how the correlation of the tidal 
fields evolves in a scale dependent way. The galaxy clusters are the best target 
for investigating the evolution of the tidal connections between large and small 
scales. Often forming at the intersections of the filaments, 
the majority of the clusters are not isolated but in constant interactions with 
the surrounding large scale structure 
\citep[e.g.,][]{millennium05}. 

The anisotropic merging of satellites from the surroundings along the major 
filaments leads to the preferential locations of the satellite galaxies near 
the minor principal axes of the external tidal fields 
(i.e., the directions of the least compression). 
As the clusters become relaxed, however, the internal tidal fields develop 
into dominance, gradually rearranging the satellite distributions and 
deviating them from the directions of the external tidal fields. 
Those satellites which merged earlier, being located in the inner regions   
of the clusters, tend to become less aligned with the directions of the 
external tidal fields.  In this picture, it may be expected that the profiles of 
the alignments between the spatial distributions of the satellites in the inner 
and the outermost regions of the clusters reflect how dominant the dark matter is
and how fast the clusters grow in the Universe   
\citep[e.g.,see][]{colberg-etal05,altay-etal06,wang-etal11,lee10,BLM11}.

Here, we determine the alignment profile of cluster galaxies using the numerical 
data from a large N-body simulation and construct a theoretical model for it. 
The organization of this paper is as follows. In \S 2, 
the numerical datasets from N-body simulations are described and the alignment 
profiles of the cluster galaxies for the three and two dimensional cases are 
determined. In \S 3, a theoretical model for the alignment profile of the 
cluster galaxies is presented and fitted to the numerical results. 
In \S 4, we discuss the possibility of using the alignment profile of cluster 
galaxies as a cosmological probe. In \S 5, the results are summarized and a final 
conclusion is reached.

\section{DATA AND ANALYSIS}

We use the semi-analytic galaxy catalogs as well as the halo catalogs\footnote{
They are available at  http://www.mpa-garching.mpg.de/millennium.} at $z=0,\ 0.5$ 
and $1$ from the Millennium Simulation for a $\Lambda$CDM concordance cosmology 
with $\Omega_{m}=0.25,\ \Omega_{\Lambda}=0.75,\ h=0.73,\ n_{s}=1.0,\ 
\sigma_{8}=0.9$ \citep{millennium05}. 
The dark matter halos and their substructures in the catalogs were found 
with the help of the standard friends-of-friends (FOF) and SUBFIND algorithms 
\citep{springel-etal01}, respectively, while the galaxies in the semianalytic 
catalogs were modeled according to the merger trees derived from the semi-
analytic simulations of the galaxy formation \citep{croton-etal06}. 

For the cluster halos with mass larger than $10^{14}\,h^{-1}M_{\odot}$ 
at each redshift, we locate their satellite galaxies from the semi-analytic 
galaxy catalog and measure the pseudo inertia tensor, $(I_{ij})$, defined in 
terms of the relative positions of the satellite galaxies located within 
cluster's virial radius $r_{vir}$ as 
\begin{equation}
\label{eqn:inertia_rv}
I_{ij}(r_{vir})=\sum_{x_{\alpha}\le r_{vir}}x_{\alpha,i}x_{\alpha,j},
\end{equation}  
where ${\bf x}_{\alpha}=(x_{\alpha,i})$ is the position vector of the $\alpha$-th 
satellite relative to the halo center and 
$x_{\alpha}\equiv \vert{\bf x}_{\alpha}\vert$ is its magnitude.
Diagonalizing $(I_{ij})$, we find the three eigenvalues and corresponding 
unit eigenvectors of the pseudo inertia tensor for each cluster halo. 
We select only those clusters which satisfy 
$\left(\varrho_{3}/\varrho_{1}\right)^{1/2}<0.9$ where $\lambda_{3}$ and 
$\lambda_{1}$ are the smallest and the largest eigenvalues of 
$I_{ij}(r_{vir})$, expecting their  pseudo inertia tensors to have three 
distinct eigenvectors. For each selected cluster halo, the major axis at the 
virial radius, ${\bf e}(r_{vir})$, is determined as the unit eigenvector 
corresponding to $\varrho_{1}$. 
 
\begin{figure}
\includegraphics[scale=1]{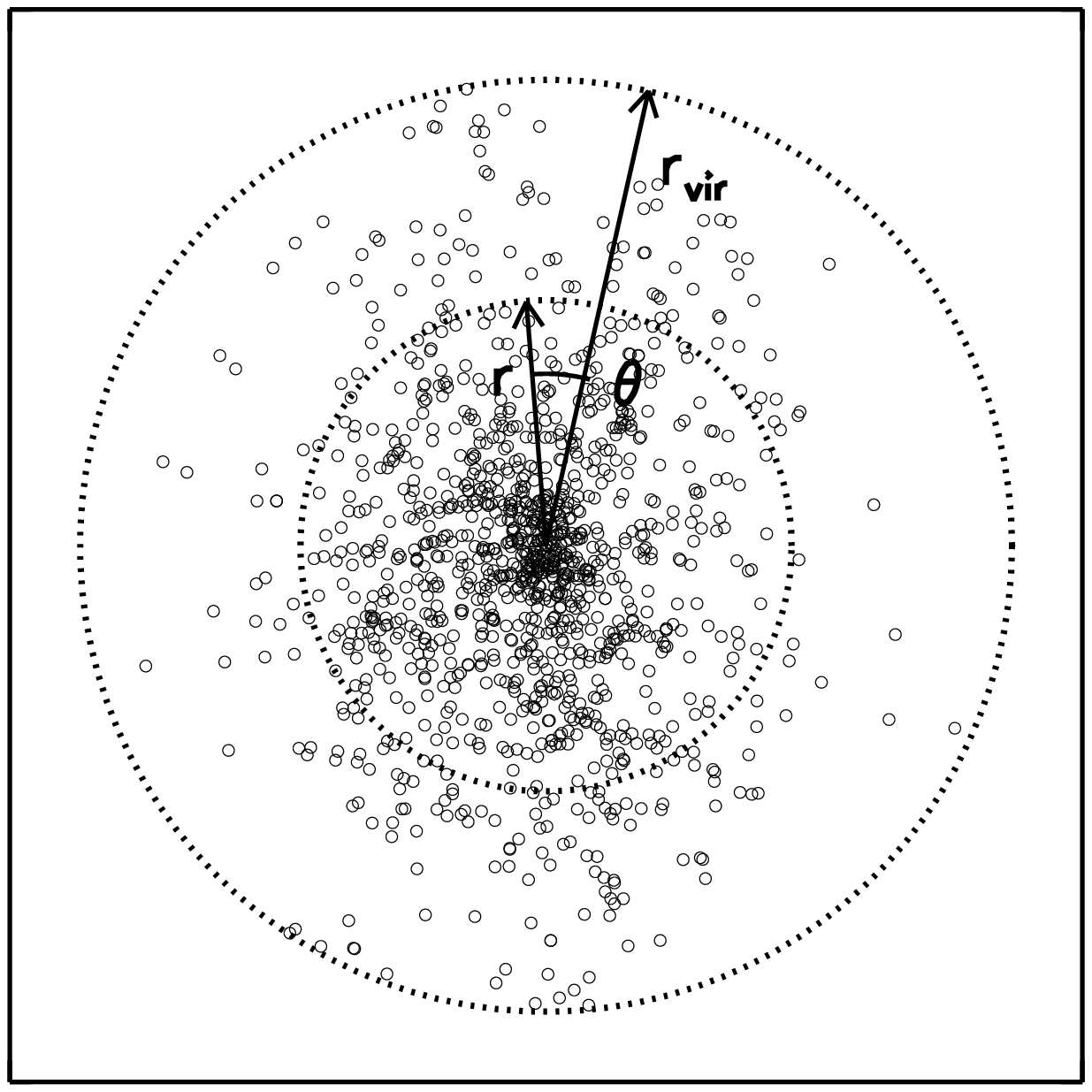}
\caption{Illustration of the anisotropic spatial distributions of the 
satellite  galaxies in the two dimensional projected plane and the angle $\theta$ 
between the major axes of the pseudo inertia tensors of the satellites within 
two different radial distances.}
\label{fig:image}
\end{figure}

Using only those satellites located within some radial distance smaller than 
the cluster's virial radius, $r<r_{vir}$, we recalculate the pseudo inertia 
tensor of each cluster's interior region measured at $r$,
\begin{equation}
\label{eqn:inertia_r}
I_{ij}(r)=\sum_{x_{\alpha}\le r}x_{\alpha,i}x_{\alpha,j}.
\end{equation}  
Diagonalizing $I_{ij}(r)$ and determining its unit eigenvector corresponding 
to the largest eigenvalue, we find the major axis, ${\bf e}(r)$,  
at the radial distance $r$, of the interior region of each cluster. 
The square of the cosine of the angle between the two major axes is now 
calculated as $\cos^{2}\theta\equiv\vert{\bf e}(r_{vir})\cdot{\bf e}(r)\vert^{2}$ 
for each selected cluster halo. Figure \ref{fig:image} illustrates the 
projected positions of the satellite galaxies in a selected cluster, 
showing how the alignment angle, $\theta$, between the distributions of the 
satellites located within two different radial distances, $r_{vir}$ and $r$ 
is measured. Varying the value of $r$ from $0.9r_{vir}$ down to some cutoff 
radial distance, $r_{c}$, defined as the radial distance within which less 
than five satellites are located,  we repeat the whole calculation. 
 
Since we are going to relate the alignment profiles of cluster galaxies to the 
auto-correlations of the tidal fields smoothed on two different top-hat 
filtering scales (see \S 3), we would like to express $\cos^{2}\theta$ as 
a function of the top-hat scale difference rather than the radial 
difference $r_{vir}-r$. For a given mass $M(\le r)$ enclosed by the 
radial distance $r$, the corresponding top-hat radial scale, 
$R$, is obtained as $R(\le r)=[3M(\le r)/(4\pi\bar{\rho})]^{1/3}$ where 
$\bar{\rho}$ is the mean mass density of the Universe.  Here, the mass, 
$M(\le r)$, can be calculated through integrating the NFW density profile 
\citep{nfw_a,nfw_b} from $0$ to $r$, as done in \citet{klypin-etal99}: 
\begin{equation}
\label{eqn:mr}
M(\le r)=M_{vir}\,\frac{f(\tilde{r})}{f(c)}, 
\end{equation} 
where $\tilde{r}\equiv r/r_{s}$ and 
$f(\tilde{r})\equiv\ln(1+\tilde{r})-\tilde{r}/(1+\tilde{r})$
Here the virial mass $M_{vir}$ is defined as 
$M_{vir}=[(4\pi/3)\rho_{c}\ 200 \ r^{3}_{vir}]$ where 
$\rho_{c}$ is the critical density at $z$ and $r_{s}$ is the scale radius, and 
$c$ is the concentration parameter defined as $c\equiv r_{vir}/r_{s}$, 
depending on $M_{vir}$ and $z$ as \citep{klypin-etal99} 
\begin{equation}
\label{eqn:c}
c\approx\frac{124}{1+z}\left(\frac{M_{vir}}{1h^{-1}M_{\odot}}\right)^{-0.084}.
\end{equation} 

From here on we express $\cos^{2}\theta$ as a function of 
$\Delta R \equiv R_{0}-R$, where $R_{0}$ and $R$ are the top-hat filtering 
scales corresponding to $M_{vir}$ at $z=0$ and $M(\le r)$, respectively, 
which can be calculated from $r_{vir}$ and $r$,  by Equations 
(\ref{eqn:mr})-(\ref{eqn:c}). The alignment profile of cluster galaxies is 
now defined as $\langle\cos^{2}\theta\rangle(\Delta R)- 1/3$,  where the constant 
factor $1/3$ is subtracted from the ensemble average since the value of the 
ensemble average will be $1/3$ when there is no alignment between the major axes 
of the pseudo inertia tensors measured on two scales, $R$ and $R_{0}$ 
\citep{pen-etal00,LP01}. 

\begin{figure}
\includegraphics[scale=0.65]{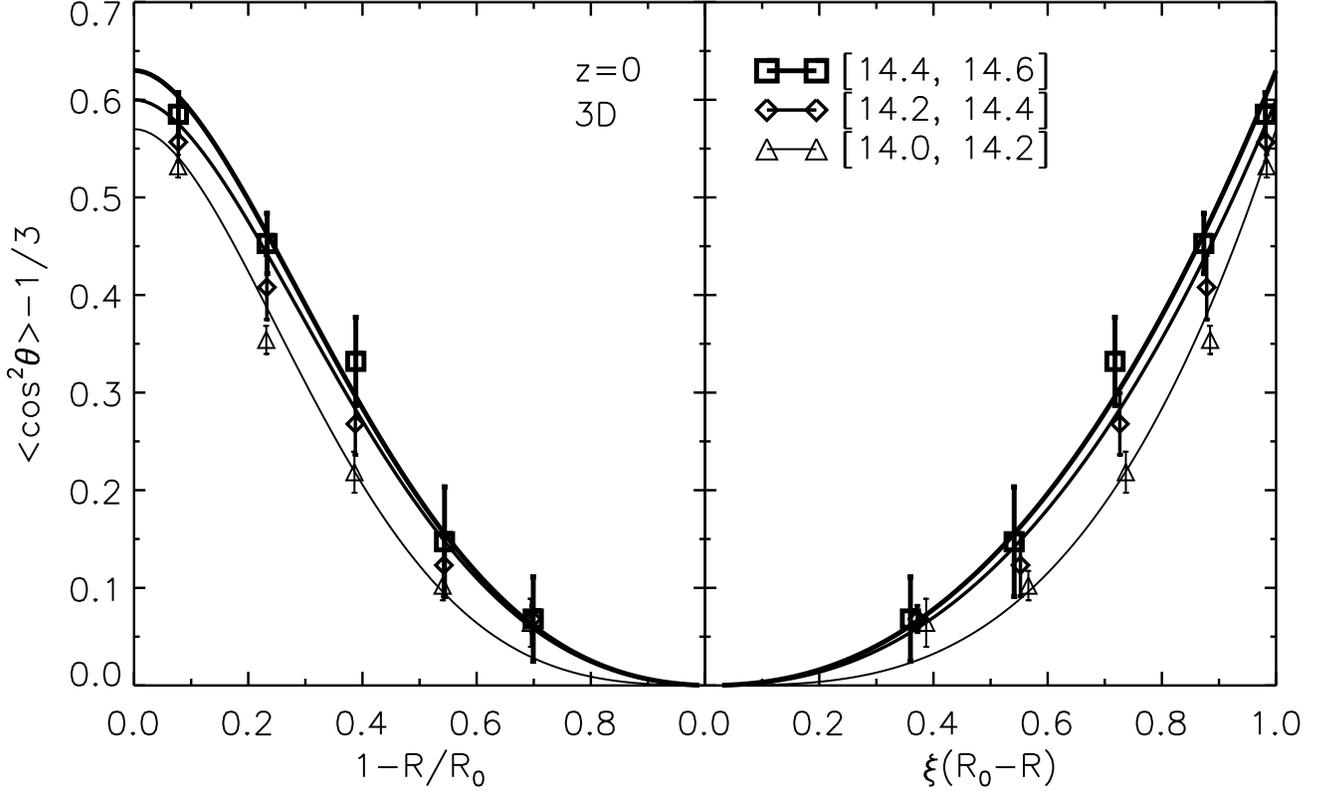}
\caption{Three dimensional alignment profiles of satellite galaxies in cluster 
halos identified at $z=0$ from the three different mass bins: 
$14.0\le \log (M/[h^{-1}M_{\odot}])\le 14.2$ (triangles); 
$14.2\le \log (M/[h^{-1}M_{\odot}])\le 14.4$ (diamonds);
$14.4\le \log (M/[h^{-1}M_{\odot}])\le 14.6$ (squares). 
The solid lines correspond to the analytic models, Equation
(\ref{eqn:profile_model}), with the best-fit parameters.}
\label{fig:profile_3dm}
\end{figure}

To investigate how the alignment profile of cluster galaxies depends on the 
mass scale, we divide the selected clusters at $z=0$ into three samples 
(I, II and III) according to their masses and separately measure the alignment 
profiles from each sample. Table \ref{tab:3d_fit} lists the redshift, logarithmic 
mass range and number of the selected clusters belonging to each sample in its 
second, third and fourth columns, respectively.  
The left panel of Figure \ref{fig:profile_3dm} plots the alignment profiles of 
cluster galaxies from the subsamples I, II, and III as open triangles, diamonds, 
and squares, respectively. 
The uncertainties associated with the numerical data points represent the 
Jackknife errors. We divide the simulation volume into eight subvolumes (the 
Jackknife resamples) each of which contains the same number of halos and then 
measure the alignment profile using the halos from each Jackknife resample, 
separately. The Jackknife errors are finally obtained as the one 
standard deviation scatter among the eight resamples \citep{WJ08}.
It is worth mentioning here that the alignment profile is plotted as a 
function of $\Delta R/R_{0}$ rather than $\Delta R$ (where $R_{0}$ is the 
top-hat filtering scale corresponding to $M_{vir}$ at $z=0$)
to make a fair comparison of the results from the three samples which must have 
different values of $R_{0}$.  As can be seen, there are clear signals of 
alignments in the small $\Delta R$ section, but the alignment profiles drop 
gradually as $\Delta R$ increases for all three cases.  The overall amplitude of 
the alignment profile shows a weak tendency to increase as the cluster's mass 
increases. 

\begin{figure}
\includegraphics[scale=0.65]{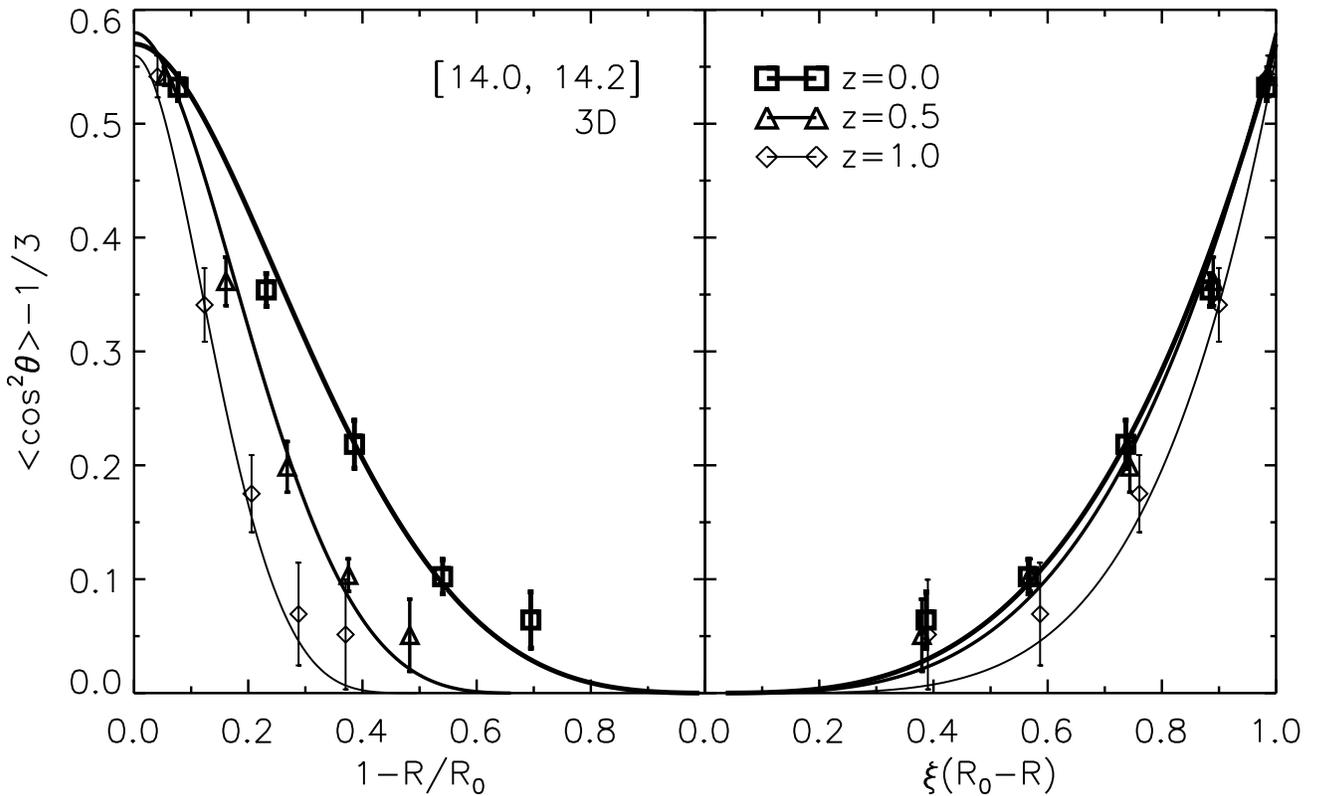}
\caption{Three dimensional alignment profiles of satellite galaxies in clusters 
with logarithmic mass in the range of $
14.0\le \log (M/[h^{-1}M_{\odot}])\le 14.2$ at three 
different redshifts: $z=0,\ 0.5,\ 1$ (open squares, triangles and diamonds, 
respectively). The solid lines correspond to the analytic models, 
Equation (\ref{eqn:profile_model}), with the best-fit parameters 
at $z=0,\ 0.5$ and $1$, respectively.}
\label{fig:profile_3dz}
\end{figure}

We also measure the alignment profiles of cluster galaxies at $z=0.5$ and $1$ 
to investigate how the profiles change with redshift. 
The left panel of Figure \ref{fig:profile_3dz} compares the three alignment 
profiles at $z=0,\ 0.5$ and $1$ (open squares, triangles and diamonds, 
respectively), where the cluster's masses lie in the narrow range of 
$10^{14}\le \log (M/[h^{-1}M_{\odot}])\le 10^{14.2}$. 
As can be seen, there is a strong tendency that the alignment profiles at higher 
redshifts drop faster with $\Delta R$, which implies that at higher redshifts the 
satellites in the inner regions of the clusters tend to be less aligned with the 
external tidal fields.  

Recalling that what we can readily observe and measure is the two 
dimensional spatial distributions of the cluster galaxies rather than the 
three dimensional ones, we also measure the alignment profiles of cluster 
galaxies in the two-dimensional projected space. Adopting the flat-sky 
approximation and taking the $z$-axis as the line-of-sight direction, 
we project the positions of all satellites onto the $x$-$y$ plane and 
follow the exactly same procedures to determine the two dimensional 
alignment profiles as $\langle\cos^{2}\theta\rangle(R-R_{0})- 1/2$ 
where the constant term $1/2$ represents the case of no alignment for 
the two dimensional case. The left panels of Figures 
\ref{fig:profile_2dm} and \ref{fig:profile_2dz} plot the same as the left panels 
of Figures \ref{fig:profile_3dm} and \ref{fig:profile_3dz}, respectively, but for 
the two dimensional case. As can be seen, the overall shapes and behaviors of the 
two dimensional alignment profiles are quite similar to the three dimensional 
ones except for the decrement of the amplitudes. 

It is worth mentioning here that we evaluate the pseudo inertia tensors for the 
derivation of the cluster galaxy alignment profile 
(Eqs.[\ref{eqn:inertia_rv}]-[\ref{eqn:inertia_r}]) rather than the conventional 
mass-weighted ones \citep[e.g.,][]{allgood-etal06}, given that 
the pseudo inertia tensors have been found to quantify well the anisotropy in 
the galaxy spatial distribution caused by the local tidal fields 
\citep[e.g.,][]{shandarin-etal06,PL07}. Since the calculation of the pseudo 
inertia tensors does not require any information on the satellite masses, the 
two dimensional alignment profile of the cluster galaxies is indeed a readily 
measurable quantity. In the following section, we provide physical explanations 
for the numerical results obtained here and present an analytic formula for the 
alignment profiles of cluster galaxies. 

\section{A THEORETICAL MODEL}

Assuming that the alignment profiles of satellite galaxies in clusters are 
directly related to the spatial correlations of the tidal fields, 
we first measure the alignments between the minor principal axes of the tidal 
fields smoothed on two different scales, using the numerical data of the 
Millennium density fields constructed on $256^{3}$ grids. As done in 
\citet{LEr07}, we perform the Fourier transform of the Millennium density 
field $\delta({\bf x})$ on each grid with the help of the FFT (Fast Fourier 
Transformation) method \citep{press-etal92} and calculate the Fourier transform 
of the tidal tensor smoothed on a given scale $R_{0}$ as 
$T_{ij}({\bf k})=k_{i}k_{j}\delta({\bf k})W(kR_{0})/k^{2}$ where 
${\bf k}\equiv (k_{i})$ is the wave vector and $W(kR_{0})$ is the top-hat window 
with the filtering radius $R_{0}$. The inverse Fourier transform of 
$T_{ij}({\bf k})$ gives the real space tidal field $T_{ij}({\bf x})$ smoothed 
on the scale of $R_{0}$ at each pixel point. The unit eigenvector of 
$T_{ij}({\bf x})$ corresponding to the smallest eigenvalue represents 
the minor principal axis,  ${\bf u}(R_{0})$. 

\begin{figure}
\includegraphics[scale=0.9]{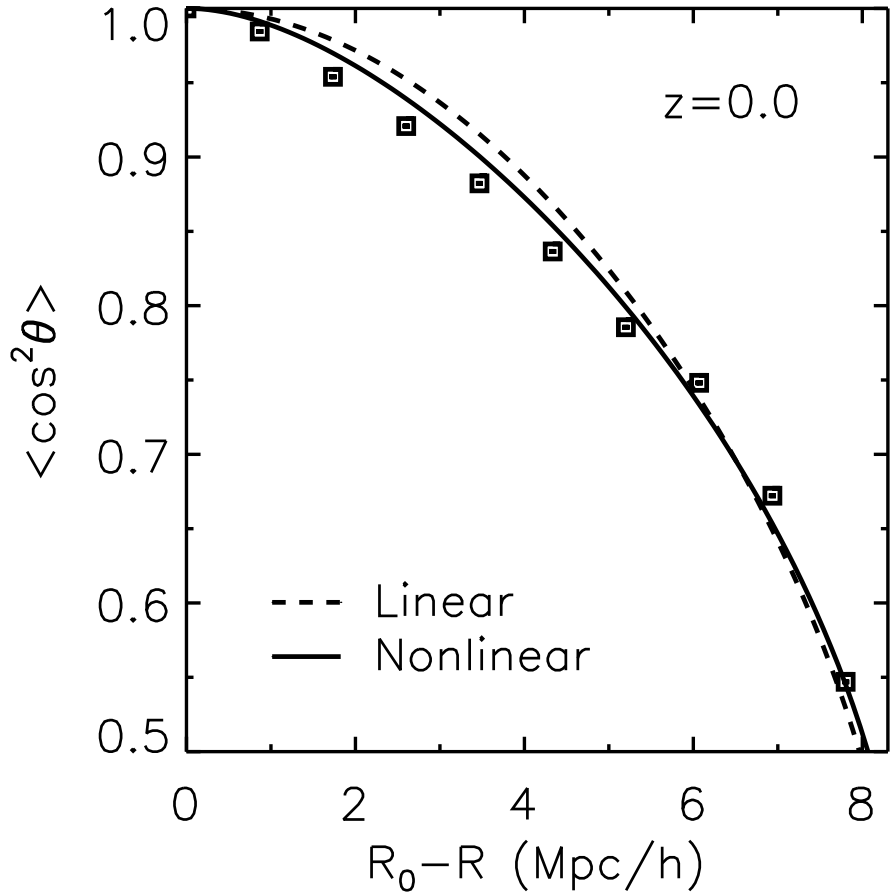}
\caption{Alignments between the minor principal axes of the Millennium tidal 
fields smoothed on two different top-hat scales (square dots) at $z=0$. 
The analytic models expressed as a power law of the density correlation coefficients
expressed in terms of linear and nonlinear density power spectrum 
(dashed and solid line, respectively).}
\label{fig:tttt}
\end{figure}

Repeating the whole process but smoothing the tidal field on some smaller scale 
$R< R_{0}$,  we also determine the minor principal axis, ${\bf u}(R)$, of the 
tidal field on scale $R$. The alignment between the minor principal axes of the 
tidal fields smoothed on two different scales, $R$ and $R_{0}$, is calculated as 
$\vert{\bf u}(R)\cdot{\bf u}(R_{0})\vert^{2}$ at each pixel. 
Varying the value of $R$, we repeat the whole calculation, 
average the alignments over $256^{3}$ pixels and express the alignment profile 
of the minor axes of the tidal fields as a function of $R_{0}-R$. Figure 
\ref{fig:tttt} plots 
$\langle\vert{\bf u}(R)\cdot{\bf u}(R_{0})\vert^{2}\rangle$ at $z=0$ 
as open squares with $R_{0}=8.7\,h^{-1}$Mpc. As can be seen, as $R$ decreases, 
the strength of the alignment gradually decreases. 

It was \citet{LP01} who have for the first time shown that the auto-correlations 
between the principal axes of the linear tidal fields smoothed on two different 
scales can be approximated as a square-root of the density correlation 
coefficient, $\xi(R_{0}-R)$, defined as \citep[see also][]{LHP09}
\begin{eqnarray}
\label{eqn:eta}
\xi(R_{0}-R) &=& 
\frac{\langle\delta_{R}\delta_{R_0}\rangle}{\sigma_{R}\sigma_{R_0}},\\
\label{eqn:dd}
\langle\delta_{R}\delta_{R_0}\rangle&\equiv&\int P(k) W(kR)W(kR_{0})
d^{3}{\bf k},\\
\label{eqn:sr}
\sigma^{2}_{R}&\equiv&\int P(k) W^{2}(kR)d^{3}{\bf k},\\
\label{eqn:sr0}
\sigma^{2}_{R_0}&\equiv&\int P(k) W^{2}(kR_{0}) d^{3}{\bf k}.
\end{eqnarray}
Adopting the model of \citet{LP01} and extending it to the nonlinear regime, 
we approximate the spatial correlations of the minor principal axes of the 
nonlinear tidal fields smoothed on the two scales of $R_{0}$ and $R$ as       
\begin{equation}
\label{eqn:tidal_model}
\langle\vert{\bf u}(R)\cdot{\bf u}(R_{0})\vert^{2}\rangle = 
\xi^{\beta}(R_{0}-R),
\end{equation}
where the power-law index, $\beta$, is an adjustable parameter. In the original 
work of \citet{LP01}, the linear power spectrum is used for the evaluation of 
the density correlation coefficient, $\xi(R_{0}-R)$, through Equations 
(\ref{eqn:eta})-(\ref{eqn:sr0}). In our work, however, we redefine 
$\xi(R_{0}-R)$ in terms of the {\it nonlinear} density power spectrum 
instead of the linear power spectrum to account for the fact that the 
alignment profiles of cluster galaxies lie in the nonlinear regime. 
For the calculation of the nonlinear power spectrum, we adopt the analytic 
formula provided by \citet{saito-etal08}. 

Adjusting the value of $\beta$,  we fit the numerical results to Equation 
(\ref{eqn:tidal_model}) via the standard $\chi^{2}$-minimization. 
The solid and dashed curves in Figure \ref{fig:tttt} represent the best-fit 
models for the cases that the nonlinear and linear density power spectrum 
are used for the calculation of $\xi$,  respectively. 
Figure \ref{fig:tttt} shows that the alignments between the minor principal 
axes of the tidal fields on different scales are indeed well approximated as 
a power-law of the density correlation coefficient especially in the large 
$\Delta R\equiv R_{0}-R$ section and that the density correlation 
coefficient expressed in terms of the nonlinear power spectrum yields a better 
fit to the numerical results. The discrepancy between the numerical results and 
the best-fit analytic curves in the small-$\Delta R$ section 
($\Delta R \le 4\,h^{-1}$Mpc) should be attributed to the low-resolution of the 
Millennium tidal fields ($256^{3}$ grids on a periodic box of 
linear size $500\,h^{-1}$Mpc). 

As the alignments between the minor principal axes of the tidal fields 
on two different scales are now found to scale as a power-law of the 
density correlation coefficient, we model the three dimensional 
alignment profile of cluster galaxies as 
\begin{equation}
\label{eqn:profile_model}
\langle\cos^{2} \theta\rangle- \frac{1}{3} = 
A\xi^{n}(\Delta R),
\end{equation}
where the amplitude $A$ and the power-law index $n$ are two adjustable 
parameters, quantifying how strongly the satellites are aligned with the 
external tidal fields at the virial radius and how fast the alignment profile 
drops with $\Delta R$, respectively. 
Adjusting the values of $A$ and $n$, we fit the numerical 
results presented in \S 2 to Equation (\ref{eqn:profile_model}) at each 
redshift. For the two dimensional case, the constant term in Equation 
(\ref{eqn:profile_model}) is changed from $1/3$ to $1/2$. For the fitting 
model, we use the density correlation coefficient, $\xi(\Delta R)$,  
expressed in terms of the nonlinear density power spectrum and set the 
value of $R_{0}$ at the maximum top-hat scale found from the clusters 
belonging to each sample.

The best-fit values of $A$ and $n$ are determined through minimizing the 
following generalized $\chi^{2}$ \citep{hartlap-etal07}:
\begin{equation}
\label{eqn:chi2}
\chi^{2} = \sum_{i,j}
\Delta\cos^{2}\theta_{i}\ C^{-1}_{ij}\ \Delta\cos^{2}\theta_{j},
\end{equation}
where $\Delta\cos^{2}\theta_{i}$ denotes the difference between the numerical 
result and the analytic model (Eq.[\ref{eqn:profile_model}]) at the $i$-th 
bin, and $(C_{ij})$ is the covariance matrix defined as 
$C_{ij}\equiv\langle(\cos^{2}\theta_{i}-\langle\cos^{2}\theta_{i}\rangle)
(\cos^{2}\theta_{j}-\langle\cos^{2}\theta_{j}\rangle)\rangle$, where the 
ensemble average is taken over the eight Jackknife resamples (see \S 2). 
The value of $R_{0}$ for the best-fit analytic model is set at the maximum 
top-hat scale calculated from the masses of the clusters 
belonging to each sample.

\begin{figure}
\includegraphics[scale=0.65]{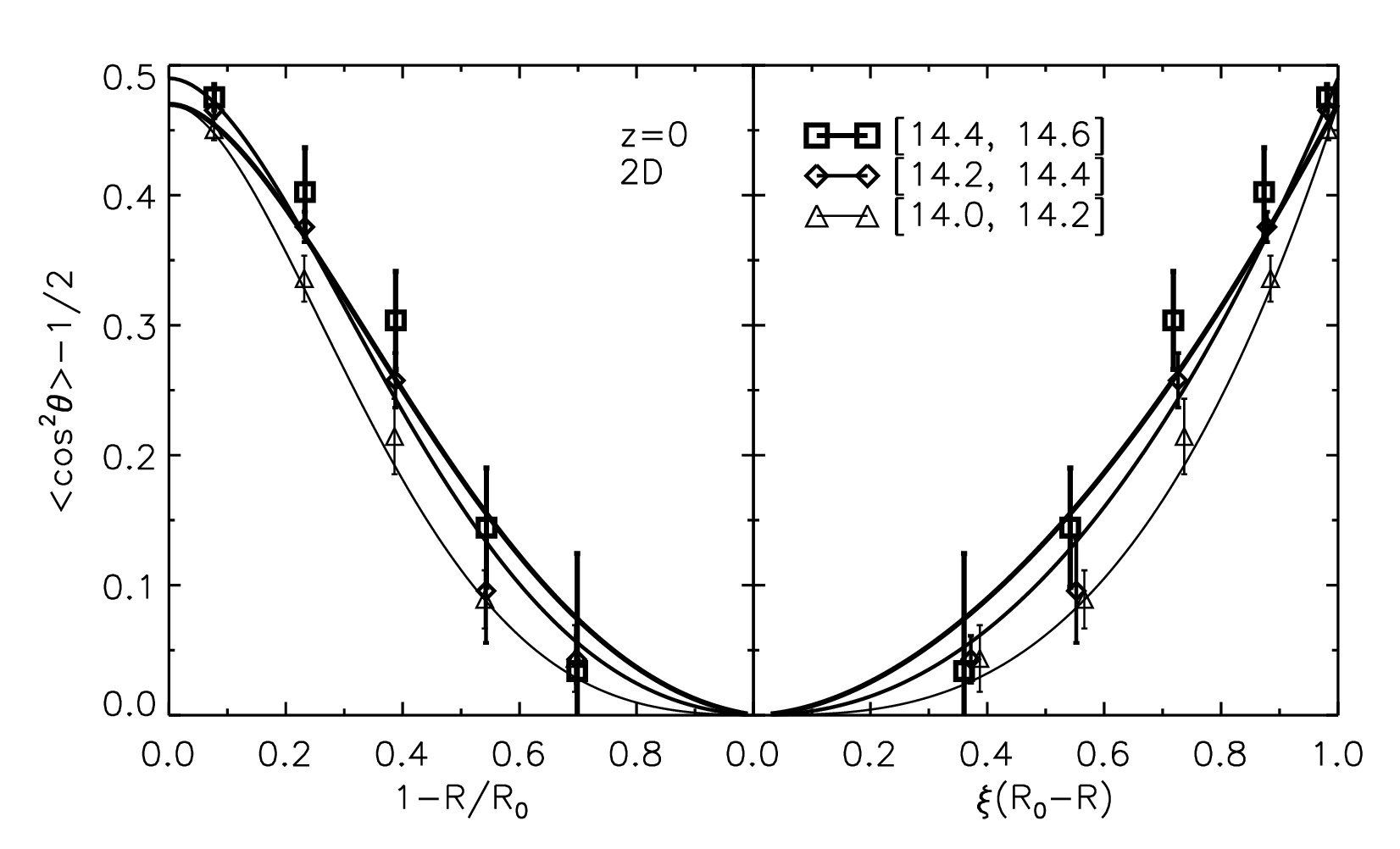}
\caption{Same as Figure \ref{fig:profile_3dm} but for the two dimensional case.}
\label{fig:profile_2dm}
\end{figure}

\begin{figure}
\includegraphics[scale=0.65]{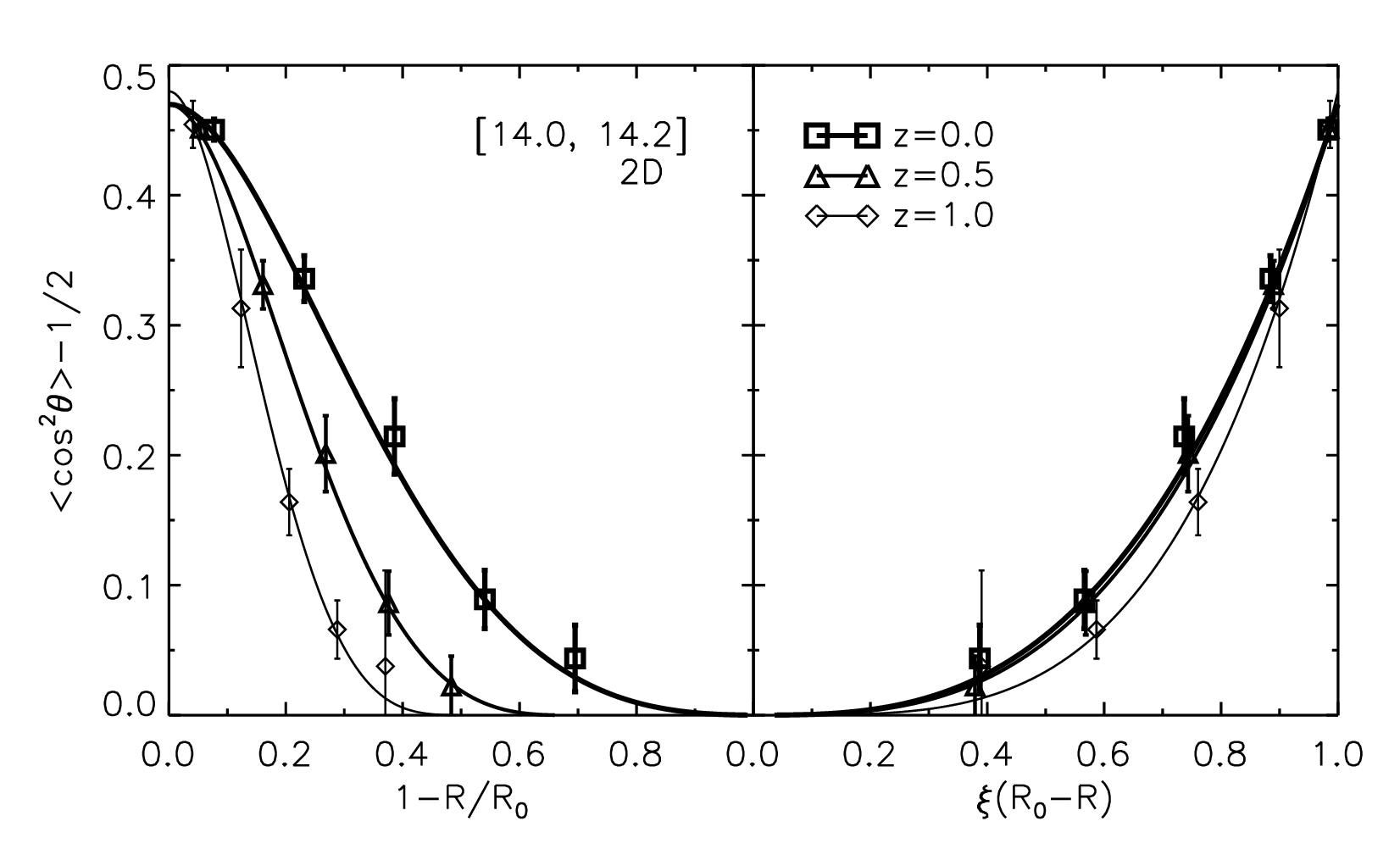}
\caption{Same as Figure \ref{fig:profile_3dz} but for the two dimensional case.}
\label{fig:profile_2dz}
\end{figure}

The best-fit amplitude and power-law index for the three and two dimensional 
cases are listed in Tables \ref{tab:3d_fit}-\ref{tab:2d_fit}. The uncertainties 
in the measurement of the best-fit parameter values represent the marginalized 
errors computed as the curvature of $\chi^{2}$ at its minimum \citep{BR96,WJ08}. 
Figures \ref{fig:profile_3dm}-\ref{fig:profile_2dz} plot the best-fit models 
as a function of $1-R/R_{0}$ in the left panels, which reveal 
that our analytic models (Eq.[\ref{eqn:profile_model}]) are indeed in good 
agreements with the numerical results for all cases. We also plot the same 
alignment profiles versus $\xi(\Delta R)$ in the right panels of Figures 
\ref{fig:profile_3dm}-\ref{fig:profile_2dz} to show explicitly that the alignment 
profiles scale as a power-law of the density correlation coefficient. 

The fitting results also show qualitatively how the alignment profiles of cluster 
galaxies change with mass scale and redshift. As the cluster mass scale 
increases, the amplitude of the alignment profile increases slightly while its 
power-law index decreases significantly. As the redshift increases, the amplitude 
of the alignment profile shows little change but its power-law index 
increases considerably. This result indicates that the satellite galaxies located 
in the massive clusters at lower redshifts tend to be more strongly 
aligned on average with the external tidal fields. To test the statistical 
significance of the dependence of the alignment profile on $\log M$ and $z$, 
we analyze a simple linear regression of the best-fit values of $A$ and $n$ 
listed in Tables \ref{tab:3d_fit}-\ref{tab:2d_fit} for the three subsamples: 
(I, II, III) and (I,IV,V), respectively. 
Table \ref{tab:alpha} which lists the slope, $\alpha$, of each linear 
model shows quantitatively that the dependences of $n$ on 
$\log M$ and $z$ are both significant with $\vert\alpha\vert> 1$ while 
the dependences of $A$ on $\log M$ and $z$ are insignificant.  

The strong dependence of $n$ on $z$ can be explained as follows.  
At higher redshifts, the relative dominance of the dark matter density parameter, 
$\Omega_{m}(z)$ is greater and thus the nonlinear effect inside the clusters 
would turn on earlier, resulting in higher degree of the deviation of the major 
axes of the pseudo inertia tensors of the cluster galaxies from the minor 
principal axes of the external tidal fields. 
As for the strong dependence of $n$ on $\log M$, the more massive clusters 
are believed to form more recently and thus their relaxation process has yet to 
be completed. In consequence, the satellite galaxies of the more massive clusters 
would retain better memory of the preferential alignments with the minor 
principal axes of the external tidal fields. 
Meanwhile, the weak dependence of $A$ on $z$ and $\log M$ may be interpreted 
as follows. The amplitudes of the alignment profiles are largely determined by 
the degree of the alignments of those satellite galaxies located at the virial 
radii of their host clusters. Since those satellite are likely to have 
just merged into the clusters, they should always exhibit the strongest 
alignments with the external tidal fields, regardless of the redshift 
and mass scale.

It is worth mentioning here that our model for the alignment profile 
of cluster galaxies does not require precisely accurate measurement of the 
cluster's virial radii, given that during the fitting procedure the value of 
$R_{0}$ for our model is set at its maximum top-hat scale from the clusters 
belonging to each sample. In other words, the alignment profile of cluster 
galaxies can be constructed even when  we have just approximate measurements 
of virial masses of the sample clusters.

\section{A COSMOLOGICAL IMPLICATION}

Although we have considered only a single cosmological model to measure the 
alignment profile of cluster galaxies that lies in the non-linear regime, the 
strong redshift dependence of the alignment profile shown in \S 3 hints that 
it might depend on the background cosmology. Furthermore, the fact that the 
alignment profile scales as a power-law of the density correlation 
coefficient, $\xi(\Delta R)$, suggests that it might be useful for breaking 
the $\Omega_{m}$-$\sigma_{8}$ degeneracy since $\xi(\Delta R)$ is independent 
of the power spectrum normalization $\sigma_{8}$. 

\begin{figure}
\includegraphics[scale=0.7]{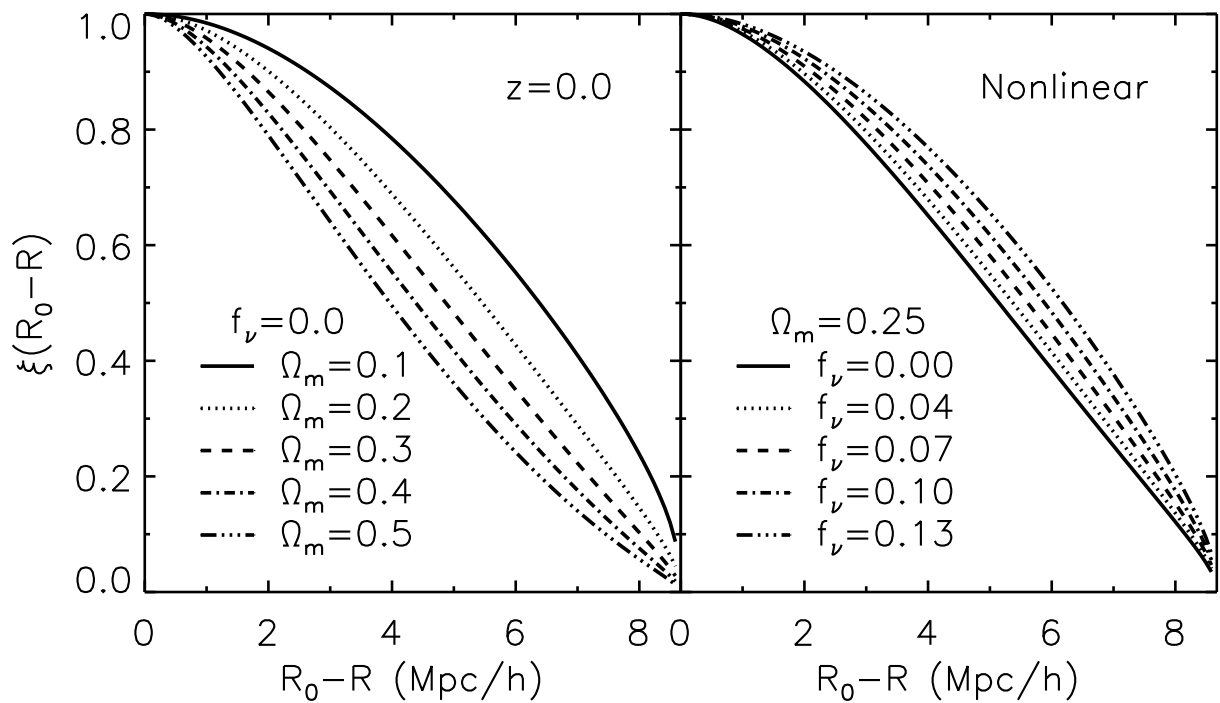}
\caption{The density correlation coefficient expressed in terms of 
the nonlinear density power spectrum for five different values of 
$\Omega_{m}$ and of $f_{\nu}$ (left and right panel, respectively). The other key 
cosmological parameters are fixed at the same values as for the Millennium 
simulations.}
\label{fig:probe}
\end{figure}

To explore this possibility, we first examine how the density correlation 
coefficient, $\xi(\Delta R)$, changes with the matter density parameter.   
Varying the value of $\Omega_{m}$ from $0.1$ to $0.5$ and fixing the other 
key cosmological parameters at the values used for the Millennium simulations, 
we repeatedly calculate $\xi(\Delta R)$ expressed in terms of the nonlinear power 
spectrum (see \S 3), at $z=0$, the results of which are plotted in the left panel 
of Figure \ref{fig:probe}. It reveals that the higher the value of $\Omega_{m}$ 
is, the more rapidly the correlation coefficient $\xi(\Delta R)$ drops with 
$\Delta R$. This result is consistent with our finding in \S 3 that the alignment 
profile of cluster galaxies drops faster with $\Delta R$ at higher redshifts when 
the relative dominance of dark matter is high. 

The other key cosmological parameter that we pay attention to is the neutrino 
mass, $m_{\nu}$.  Recently, \citet{neutrino11} have demonstrated by hydrodynamic 
simulations that if the neutrinos have non-zero mass (in a $\Lambda$CDM+$\nu$ 
cosmology), then their free-streaming would have a strong impact on the formation 
and evolution of the large-scale structure, suppressing the formation of small-
scale objects and thus decreasing the abundance of galaxy clusters and their 
clustering strength \citep[see][for a recent review]{LP06}. 
Yet, they emphasized that these effect caused by the neutrino free streaming 
are strongly degenerate with the power spectrum normalization factor, 
$\sigma_{8}$.  

It is expected that if there are larger amount of free-streaming neutrinos, 
then they would defer the merging of the satellite galaxies into 
the clusters and thus the cluster galaxies would keep better the memory of the 
external tidal fields. The higher the fraction of massive neutrinos is, the 
higher amplitude and lower power-law index the alignment profile of cluster 
galaxies has. Under the speculation that our analytic model for a $\Lambda$CDM 
cosmology can be extended to $\Lambda+\nu$CDM cosmology,   
we calculate $\xi(\Delta R)$ for five different cases of the neutrino mass 
fraction, $f_{\nu}\equiv \Omega_{\nu}/\Omega_{m}$. Regarding the nonlinear matter 
density power spectrum for a $\Lambda$CDM+$\nu$ cosmology, we use the 
analytic formula provided by \citet{saito-etal08}.
The right panel of Figure \ref{fig:probe} plots the matter density correlation 
coefficient at $z=0$ for five different cases of $f_{\nu}$. The higher the 
value of $f_{\nu}$ is, the correlation coefficient, $\xi(\Delta R)$, drops 
less rapidly with $\Delta R$, consistent with our expectation. 

\section{SUMMARY AND CONCLUSION}

We have determined the three and two dimensional alignment profiles of cluster 
galaxies using the halo and semi-analytic galaxy catalogs from the Millennium 
simulation.  The alignment profiles of cluster galaxies are defined as the 
average alignments between the major axes of the pseudo inertia tensors of 
the cluster galaxies on two different top-hat scales $R_{0}$ and $R$ 
(where $R_{0}$ is the top-hat radius enclosing cluster's virial mass and 
$R<R_{0}$). The alignment profiles of cluster galaxies reflect how well 
the satellite galaxies, after merging into the clusters, retain the memory of 
the preferential alignments with the minor principal axes of the external 
tidal fields and how fast they gradually lose the alignment tendency 
under the influence of the internal tidal fields as the cluster's relaxation 
proceeds and nonlinear effect becomes stronger.

It has been shown that the alignment profiles of cluster galaxies expressed as a 
function of $\Delta R=R_{0}-R$ behave as a power-law of the density correlation 
coefficient. The alignment profiles exhibit weak dependence of the amplitude $A$
on mass scale and redshift but strong dependence of the power-law index $n$
whose statistical significance is tested by using a simple linear regression analysis.
As the cluster mass decreases and the redshift increases, 
the alignment profile drops faster with $\Delta R$, 
which is consistent with the scenario that the faster merging process of the 
satellites and earlier onset of the nonlinear effect inside the clusters tend to 
break more easily the preferential alignments of the cluster galaxies with the 
minor principal axes of the external tidal fields. The weak dependence of $A$
on mass scale and redshift implies that the satellite galaxies 
located at the virial radii of their host clusters always keep good memory of the 
external tidal fields regardless of the mass scale and redshift.

We have suggested that the alignment profiles of cluster galaxies might be 
a complimentary cosmological probe, being useful especially for breaking 
the $\Omega_{m}$-$\sigma_{8}$ and $f_{\nu}$-$\sigma_{8}$ degeneracies 
(where $f_{\nu}$ is the neutrino mass fraction) owing to the 
insensitivity of $\xi(\Delta R)$ to $\sigma_{8}$.  
To prove this theoretical concept, however, it will be necessary to 
test the validity of our model against N-body simulations for different 
cosmologies other than the $\Lambda$CDM model assumed for the Millennium 
simulations. As a large catalog of the galaxy clusters and their 
satellites from the Sloan Digital Sky Survey Data Release 7 is in the pipeline 
\citep{hao-etal10}, our future work is in the direction of conducting numerical 
and observational tests of the alignment profiles of cluster galaxies 
presented here.

\acknowledgments

We thank an anonymous referee for the very useful comments and suggestions. 
The Millennium Simulation analyzed in this paper was carried out by the 
Virgo Supercomputing Consortium at the Computing Center of the Max-Planck Society 
in Garching, Germany. The simulation databases and the web application providing 
online access to them were constructed as part of the activities of the German 
Astrophysical Virtual Observatory. This work was supported by the National 
Research Foundation of Korea (NRF) grant funded by the Korea government (MEST, 
No.2011-0007819). Support for this work was also provided by the National 
Research Foundation of Korea to the Center for Galaxy Evolution Research.


\clearpage
\clearpage
\clearpage
\clearpage
\clearpage
\clearpage
\clearpage
\clearpage
\begin{deluxetable}{cccccc}
\tablewidth{0pt}
\setlength{\tabcolsep}{5mm}
\tablecaption{Redshift, logarithmic mass range, number of the selected 
clusters, best-fit amplitude, and best-fit power-law index of the 
three dimensional alignment profiles of satellite galaxies in the clusters 
from the five samples}
\tablehead{sample & $z$ & $\log(M/[h^{-1}M_{\odot}])$ 
&  $N_{c}$ & $A$ & $n$} 
\startdata
I   & $0$   & $[14.0,\ 14.2]$ & $1250$ & $0.570\pm 0.006$ & $3.130\pm 0.038$\\
II  & $0$   & $[14.2,\ 14.4]$ & $649$  & $0.600\pm 0.010$ & $2.350\pm 0.017$\\
III & $0$   & $[14.4,\ 14.6]$ & $337$  & $0.630\pm 0.006$ & $2.280\pm 0.090$\\
IV  & $0.5$ & $[14.0,\ 14.2]$ & $832$  & $0.580\pm 0.005$ & $3.420\pm 0.041$\\ 
V   & $1$   & $[14.0,\ 14.2]$ & $369$  & $0.560\pm 0.005$ & $4.700\pm 0.042$
\enddata
\label{tab:3d_fit}
\end{deluxetable}
\clearpage
\begin{deluxetable}{cccc}
\tablewidth{0pt}
\setlength{\tabcolsep}{5mm}
\tablecaption{Same as Table \ref{tab:3d_fit} but for the two dimensional 
alignment profiles}
\tablehead{sample & $N_{c}$ & $A$ & $n$} 
\startdata
I   & $1077$ & $0.470\pm0.004$ & $2.930\pm0.049$\\
II  & $552$  & $0.490\pm0.001$ & $2.190\pm0.021$\\
III & $289$  & $0.470\pm0.004$ & $1.810\pm0.096$\\
IV  & $724$  & $0.470\pm0.003$ & $3.060\pm0.065$\\ 
V   & $324$  & $0.480\pm0.005$ & $3.840\pm0.050$
\enddata
\label{tab:2d_fit}
\end{deluxetable}
\clearpage
\begin{deluxetable}{ccc}
\tablewidth{0pt}
\setlength{\tabcolsep}{5mm}
\tablecaption{Result of the linear regression test of the dependence of the 
best-fit parameters of the alignment profiles on the logarithmic mass and 
redshift}
\tablehead{function & $\alpha$} 
\startdata
$A(\log M)$ & $+0.100$\\
$n(\log M)$ & $-1.417$\\
$A(z)$    & $-0.007$\\
$n(z)$    & $+1.047$
\enddata
\label{tab:alpha}
\end{deluxetable}

\end{document}